\documentclass[twocolumn,english,preamssymb,fleqn,10pt]{revtex4}  
\usepackage{epsfig,amssymb,amsmath,graphicx,subfigure,hyperref,float}   
\usepackage{color}

\newcommand{\be}{\begin{eqnarray}}  
\newcommand{\ee}{\end{eqnarray}}

\begin{document}
\title{Fractionalization of Interstitials in Curved Colloidal Crystals}
\author{William T.M. Irvine$^{1}$, Mark J. Bowick$^2$ and Paul M. Chaikin$^3$}
\affiliation{$^1$ James Franck Institute, University of Chicago, 929 E 57$^{\rm th}$ street, Chicago, IL 60637, USA\\
$^2$  Physics Department, Syracuse University, Syracuse NY 13244-1130, USA\\
$^3$ Center for Soft Matter Research, New York University, 4 Washington Place, New York, NY 10003, USA
} 

\begin{abstract}
Understanding the out-of equilibrium behaviour of point defects in crystals, yields insights into the nature and fragility of the ordered state, as well as being of great practical importance. In some rare cases defects are spontaneously healed -  a {\it one}-dimensional crystal formed by a line of  identical charged particles, for example, can accommodate an interstitial (extra particle) by a re-adjusting all particle positions to even out the spacing.  In sharp contrast, particles organized into a perfect hexagonal crystal in the plane (Fig.~1E) cannot accommodate an interstitial by a simple re-adjustment of the particle spacing -  the interstitial remains instead trapped between lattice sites and diffuses by hopping\cite{PL:2001,PL:2001b,PL:2001c}, leaving the crystal permanently defected.  Here we report on the behavior of interstitials in colloidal crystals on {\em curved} surfaces (Fig.~1A,B). Using optical tweezers operated independently of three dimensional imaging, we insert a colloidal interstitial in a lattice of similar particles on flat and curved (positively and negatively) oil-glycerol interfaces and image the ensuing dynamics.  We find that, unlike in flat space, the curved crystals self-heal through a collective rearrangement that re-distributes the increased density associated with the interstitial. The self-healing process can be interpreted in terms of an out of equilibrium interaction of topological defects with each other and with the underlying curvature.  Our observations suggest the existence of ``particle fractionalization'' on curved surface crystals.
 \end{abstract}
 
\maketitle

Much in the way that  a sticker does not naturally fit on the surface of a car bumper, crystals must deform to fit on curved surfaces.  Curvature changes the rules of geometry.   The interior angles in a triangle, for example, no longer add to 180$^\circ$ and initially parallel lines can diverge or converge leading to compression and stretching.  This very general  geometry-induced frustration applies to   any phase that possesses orientational order in flat space such as nematics, smectics and crystals.

A recent flurry of activity has investigated how a crystal can undergo structural changes to relieve this frustration by introducing topological defects. In hexagonal crystals, two types of topological defect are found: disclinations (Fig.~1F,G), that correspond to a missing (inserted) $60^{\circ}$ wedge of lattice and disrupt orientational order, and dislocations (Fig.~1H), that correspond to two extra rows of particles that terminate at the core of the defect and disrupt translational order.    Clearly both are non-local in origin and influence\cite{NelsonBook} and disrupt order in the crystal.
On surfaces that are  {\it curved}, however, topological defects can play a complementary role, relieving both compressive and shear stress\cite{nelsonpeliti,PDM,BNT:2000,vv,BB2003,IVC:2010}.  For example on a sphere, on which the sum of the interior angles in a triangle is increased from $180^{\circ}$ by the Gaussian curvature, 5-fold coordinated disclinations can make up this angular excess (as seen on a soccer ball with its twelve pentagonal panels). When the lattice spacing is much less than the radius of curvature of the surface, the isotropic and shear stress created by curvature is relieved by groups of defects that organize into grain-boundary-like structures that  freely terminate within the medium, characterized by the excess disclination charge; examples of structures both  neutral (`pleats'\cite{IVC:2010}) and charged (`scars'\cite{BNT:2000})   are  shown in Figs 1C and 1D.

\begin{figure}
\centering
\includegraphics[width=\columnwidth]{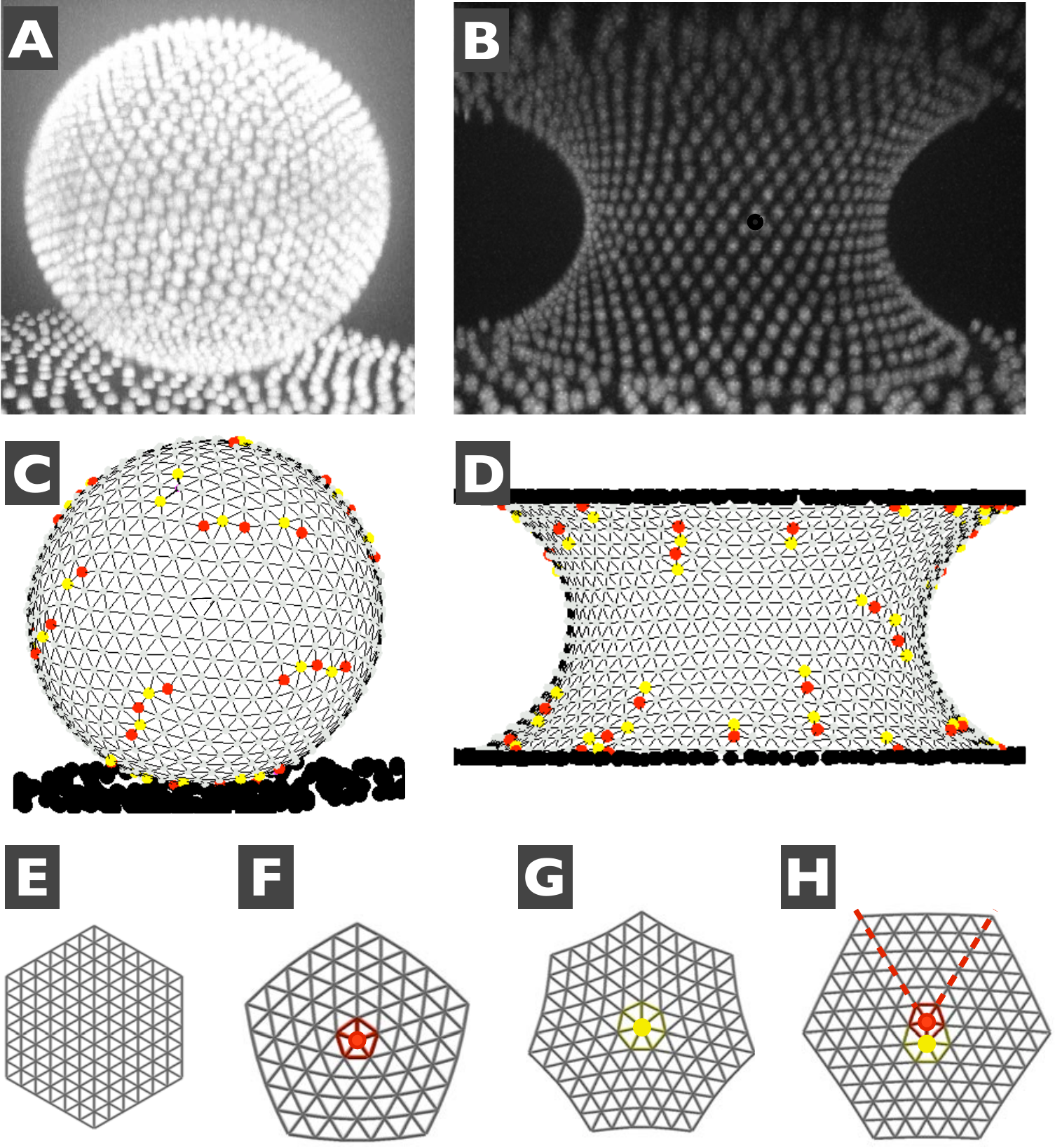}
\caption{Curved crystals and topological defects.
(A) A spherical crystal formed by self-assembled colloidal beads on a liquid droplet (Diameter $\sim$60$\mu m$). 
(B) A negative curvature crystal formed by the same colloidal beads on the surface of a capillary bridge (Bridge diameter $\sim$45$\mu m$).
(C) A Delaunay triangulation of a typical equilibrated configuration for the crystal (A) with 5-coordinated particles shown in red and 7-coordinated particles shown in yellow.
(D) A Delaunay triangulation of a typical equilibrated configuration for the crystal (B).
(E) A regular hexagonal lattice configuration.
(F) A 5-disclination in a planar crystal. 
(G) A 7-disclination in a planar crystal. 
(H) A dislocation in a planar crystal.}
\label{spherical_xtal}
\end{figure}

Our experimental system consists of positively charged colloidal PMMA particles ($ \sim 2 \mu$m in  diameter) coated in a layer of (poly)hydroxy stearic acid and suspended in oil (a CHB/dodecane mixture). In the presence of an oil-glycerol interface, image charge effects drive the binding of the particles to the interface,  where, without wetting, they form a monolayer (Fig.~2A). The repulsive screened Coulomb interactions cause them to self-organize into a crystalline lattice that conforms to the curved surface. By index matching the mixture of CHB and dodecane to the glycerol  we obtain a clean system that can be imaged fully using a confocal microscope (Fig.~2C). The system equilibrates into scarred and pleated structures\cite{IVC:2010}. 

To study the  behavior of interstitials we add a particle to this system using optical tweezers decoupled from the imaging (see methods) and, by simultaneously imaging in three dimensions, we follow the out-of-equilibrium dynamics of the defects on the curved surface. 

Interstitial defects have a  local material origin, resulting from  a  {\em single} extra particle  forced into an otherwise ordered crystalline array. 
When a particle that is identical to the  other particles  is added to the crystal, however, its identity becomes ambiguous;   the crystal accommodates its presence by  local re-adjustments that leave  two signatures of the interstitial's presence. The first is a localized spike in the compressional strain (density) field that corresponds to the additional particle's mass; the second is  a bound triplet or doublet of dislocations\cite{DBR:1979,EV:1991,SDR:2000} as can be seen in Fig.~2B and supplementary movie 2.  In flat space these dislocations remain bound, affecting neither translational nor orientational order.

\begin{figure}
\centering
\includegraphics[scale=0.35]{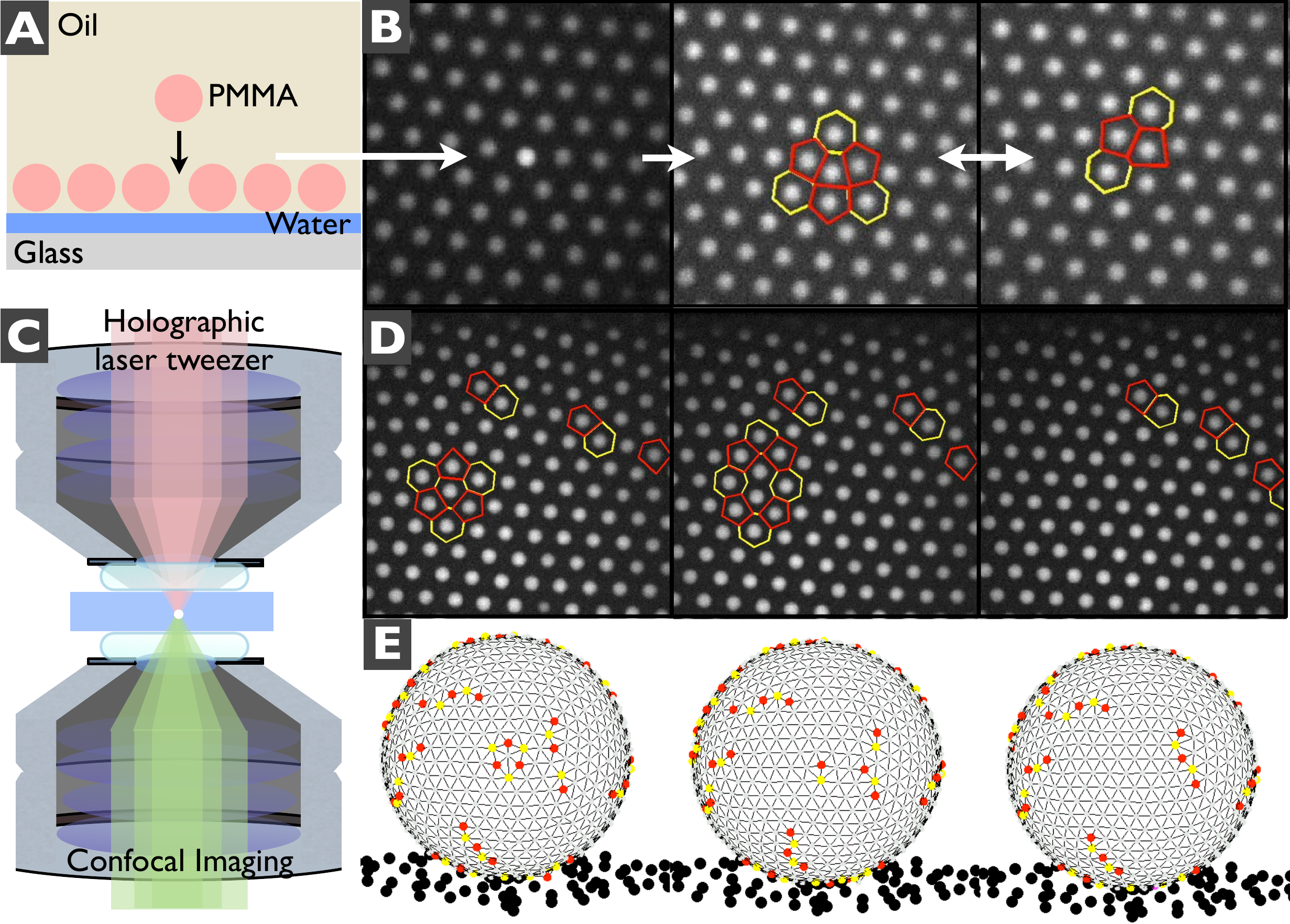}
\caption{Interstitials in flat space and interstitial absorption by grain boundaries. 
(A) Inserting a particle to create a self-interstitial.
(B) An interstitial in flat space typically evolves into bound states of three or two dislocations with vanishing net Burgers' vector (lattice spacting $\sim3\mu m$).
(C) Schematic of confocal imaging combined with laser tweezers.
(D) An interstitial in flat space close to a grain boundary is absorbed by the latter (See supplementary movies).
(E) An interstitial created on a spherical crystal and subsequently absorbed by a grain boundary scar.} \label{inter_absor}
\end{figure}

\begin{figure*}[t]
\centering
\includegraphics[width=2 \columnwidth]{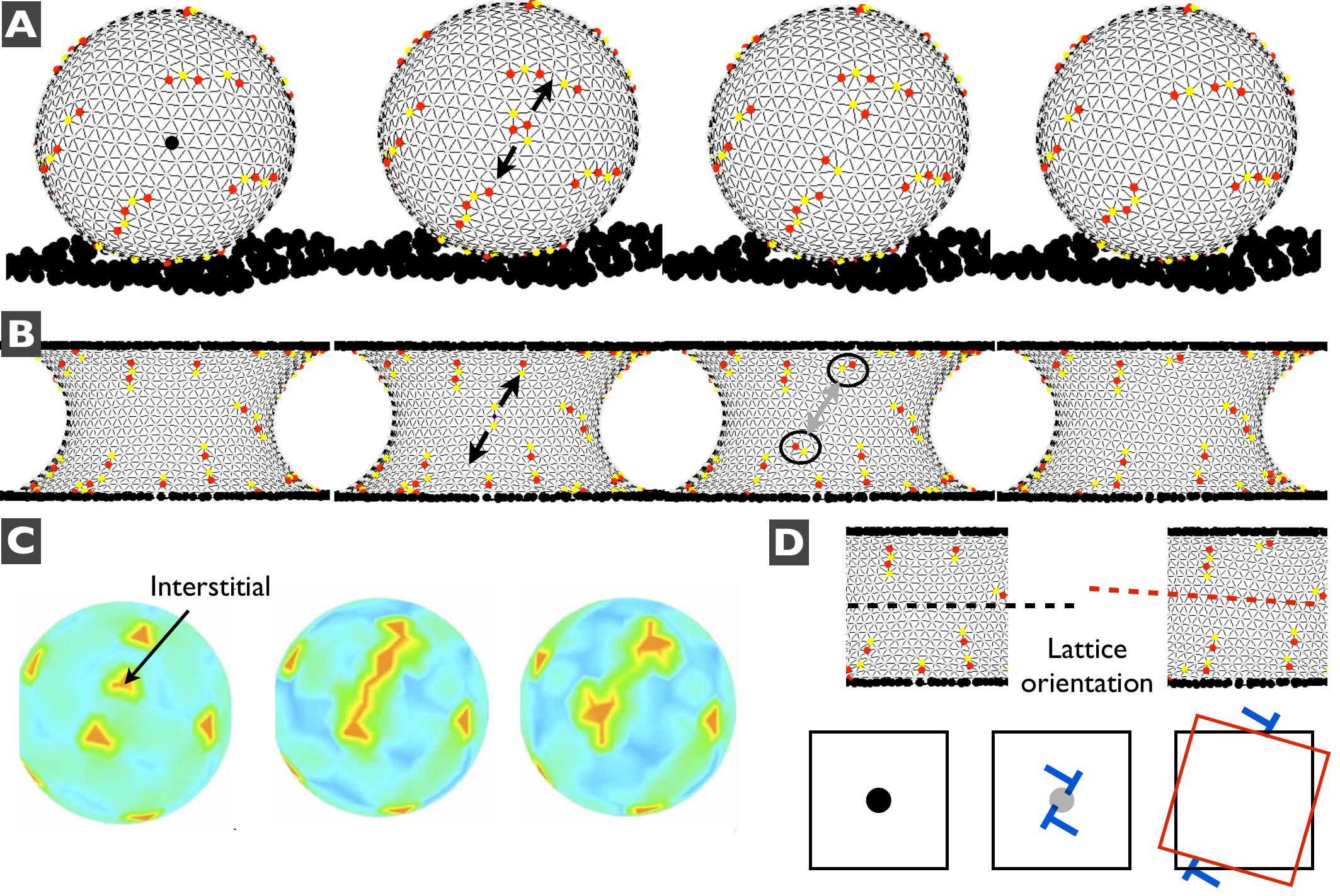}
\caption{Interstitial fractionalization.
(A) The insertion of an interstitial in a spherical crystal is followed by its fissioning into two dislocations which migrate, gliding  along parallel Bragg planes in opposite directions, and are 
subsequently absorbed into existing grain boundary scars. 
(B) The insertion of an interstitial on a negatively curved capillary bridge followed by its fissioning into two dislocations - the upper dislocation is absorbed by an 
existing dislocation which rotates to absorb the Burgers' vector and the bottom dislocation is absorbed by an existing grain boundary scar.
(C) Plot of the compressional strain distribution that accompanies a fractionalization event as evaluated numerically on a sphere.
(D) The twisting of Bragg rows resulting from interstitial fractionalization as deduced from imaged configurations before and after fractionalization; the schematic illustrates the twisting of the crystalline patch containing an interstitial as it constituent dislocations separate to globally distinct locations on the lattice.\label{inter_frac}
} 
\end{figure*}

Once added, an interstitial can move or diffuse  by hopping between lattice inter-sites\cite{PL:2001,PL:2001b,PL:2001c} with the dislocations remaining bound, preserving the local character of the interstitial. Note that the original particle need not diffuse with the interstitial defect, but rather may remain in the region it was added, as can be verified by tracking the added particle in this case.

The diffusive motion can be biased by  stress fields\cite{HL:1968}. We observe this in both flat and curved space: If the particle is added in close proximity to a grain boundary, the latter can absorb the interstitial with little change in structure, while eliminating the stress energy associated with the interstitial. The interstitial  hops towards the grain boundary and is absorbed. This is the case in flat or curved space alike - as can be seen in Fig.~2D,E and supplementary movies 3 and 4.

Adding particles to crystals bound to positively and negatively curved surfaces, however,  we  observe an additional, strikingly  different, behavior: the addition of a particle  is followed by the fissioning of the normally bound dislocations into pairs that travel, gliding along parallel Bragg planes in opposite directions, leaving the crystal region in which the particle was added rotated (Fig.~3A for a spherical crystal and  in Fig.~3B,D for a negatively curved crystal, see also supplementary movies 6 and 7). 

We find that this intriguing mechanism, predicted theoretically and numerically in the special case of un-equilibrated arrangements of disclinations on spheres\cite{BNS:2007,BST:2007}, occurs generically in experiments on equilibrated spherical droplets and equilibrated negatively curved capillary bridges.  

This non-local  behavior raises the question - where does the particle go?   By partially bleaching the particles and subsequently adding an unbleached, and therefore brighter particle to the crystal (Fig.~1B), we verified that in both types of instability, the specific particle that was added  remains in the region in which it was added,  in agreement with the behavior in flat space. However in the case where the dislocations remain bound as they diffuse in flat space, or migrate towards a grain boundary as seen before, the density increase associated with the added particle remains localized  and has a continuous trajectory to the grain boundary. In contrast, for the case in which the interstitial fractionalizes,  the mass transport cannot be similarly localized - though the mass associated with an extra particle clearly  leaves the region. 

We investigated this transport through a numerical (Fig.~3C) and experimental (Supplementary movie 3) investigation of the stress fields associated with a fractionalization event.  Fig.~3C shows how the compressive part of the stress field in the case of the sphere extends, creating two branches that join up to the scars, effectively delocalizing the increased density associated with the particle in the process.

While fractionalization could also be achieved in planar crystals by applying a strong shear to a region enclosing an interstitial and could in principle occur if a particle is added in the center of a small grain, these represent very special conditions in flat space,   whereas the conditions occur generically in curved space.
Since defects have localized cores with quantized angular charge, the angular frustration generated by curvature is more gradual and some residual stress remains in the regions in between pleats or scars. It is this residual stress that drives the self-healing.

Just as topological defects find a new life on curved surfaces, going from order-disrupting excitations to order-restoring,   interstitials that are normally localized stable point defects develop complementary non-local character.    The spontaneous morphing of  interstitials into sets of dislocations gliding through the lattice provides particle arrays access to configurations that might normally be difficult to reach without additional thermal noise or external perturbations.  This allows new mechanisms for the equilibration of ordered phases on curved space which could prove important for the practical realization  of self assembly schemes based on defects\cite{Stellacci}. Our unique  experimental  combination of a versatile model system with  with  full three dimensional  control and imaging  further paves the way for studying non-equilibrium effects in ordered and disordered phases in arbitrarily curved interfaces. 

\noindent{\bf Methods}

\noindent The PMMA particles, synthesized following the methods of Refs.\cite{antl,bosma} are coated with a layer of poly(hydroxy stearic acid) (PHSA), which charges positively ($\sim$100 charges per particle) in the oil\cite{mirj}. The particles were suspended in mixture of Cyclohexyl-Bromide and Dodecane (approx. 72\% w/w) that was mixed before each experimental run to match the refractive index of Glycerol as measured by an Abbe refractometer. This avoids lensing by the oil-glycerol interface, while allowing for a small index contrast between the particles and the oil phase for optical tweezing. The Glycerol-Oil interfaces were prepared by emulsification (in the case of spherical surfaces) and by deposition of glycerol droplets in contact with air in capillary channels that were  subsequently filled by the particle suspension (in the case of capillary bridges) and sealed to avoid evaporation. The samples were imaged using a Yokogawa CSU-10 spinning disk confocal. Optical tweezing independent of the confocal imaging was achieved by substituting the microscope condenser with a second microscope objective. A holographically shaped 1064nm trapping laser was then projected through the upper objective into the sample. Particle location was determined from the images using the IDL routines of Ref.\cite{crock} and triangulation and defect identification using custom codes written in Matlab. 

\noindent{\bf Acknowledgements}

\noindent We acknowledge discussions with S.~Sacanna, A.D.~Hollingsworth, A.~Grosberg, T.~Witten and V.~Vitelli.  This work was supported by Rhodia and the English speaking union (WTMI), the National Science Foundation grant DMR-0808812 (MJB), the MRSEC Program of the National Science Foundation under Award Number DMR-0820341 and NASA NNX08AK04G (PMC).

WTMI and MJB acknowledge hospitality from the Aspen Center for Physics. Correspondence and requests for materials should be addressed to WTMI (email:wtmirvine@uchicago.edu), PMC (email:pc86@nyu.edu) and MJB (email: bowick@physics.syr.edu).

\end{document}